\documentclass[aps,pra,twocolumn,superscriptaddress,showpacs]{revtex4-1}

\usepackage{graphicx}
\usepackage{amsmath}
\usepackage{cleveref}
\usepackage{color}

\usepackage[bookmarks=true,colorlinks=true,urlcolor=blue,linkcolor=blue,citecolor=blue,breaklinks]{hyperref}

\usepackage{amssymb}
\usepackage{booktabs}

\begin{document}

\title{Nonadiabatic coupling effects in MgB\texorpdfstring{$_2$}{TEXT} reexamined}

\newcommand*{\DIPC}[0]{{
Donostia International Physics Center (DIPC),
Paseo Manuel de Lardizabal 4, 20018 Donostia-San Sebasti\'an, Spain}}

\newcommand*{\FUBER}{{
Institut f{\"u}r Chemie und Biochemie, Freie Universit{\"a}t Berlin,
Takustr.~3, 14195 Berlin, Germany}}

\newcommand*{\IFS}[0]{{
Center of Excellence for Advanced Materials and Sensing Devices, Institute of Physics, Bijeni\v{c}ka 46,
10000 Zagreb, Croatia}}

\author{Dino Novko}
\email{dino.novko@gmail.com}
\affiliation{\IFS}
\affiliation{\DIPC}

\date{\today}

\begin{abstract}
The unusual Raman spectrum of MgB$_2$ and its formidable temperature dependence are successfully reproduced by means of a parameter-free \emph{ab initio} nonadiabatic theory that accounts for the electron-hole pair scattering mechanisms with the system phonons. This example turns out to be a prototypical case where a strong nonadiabatic renormalization of the phonon frequency is partially washed out by the aforementioned scattering events, bringing along a characteristic temperature dependence. Both electron-hole pair lifetime and energy renormalization effects due to dynamical electron-phonon coupling turn out to play a crucial role. This theory could aid in comprehending other Raman spectra characterized with unconventionally strong electron-phonon interaction.
\end{abstract}


\maketitle




Recent years have witnessed an ample interest in nonadiabatic (NA) coupling effects\,\cite{bib:brovman67,bib:geilikman71} and their intriguing impact on the vibrational properties of solids\,\cite{bib:giustino17}. Particularly strong deviations from the adiabatic phonon spectrum were observed for the long wavelength ($\mathbf{q\approx 0}$) modes of carbon-based materials, such as metallic carbon nanotubes\,\cite{bib:caudal07,bib:piscanec07}, graphite intercalation compounds\,\cite{bib:saitta08,bib:zhao11,bib:chacontorres12,bib:chacontorres13}, graphene\,\cite{bib:lazzeri06,bib:pisana07,bib:howard11}, and boron-doped diamond\,\cite{bib:caruso17}. The corresponding NA theory based in first principles is well established and in most instances the calculated NA phonon frequencies complement the experiments quite accurately\,\cite{bib:lazzeri06,bib:saitta08,bib:calandra10}.
%
%
Remarkably, in MgB$_2$, both the state-of-the-art adiabatic and NA descriptions break down. Namely, the Raman measurements reveal an unusually large linewidth of the $E_{2g}$ phonon peaked at around 77\,meV\,\cite{bib:quilty02,bib:rafailov02,bib:martinho03,bib:shi04,bib:ponosov17}, which falls just between the adiabatic (67\,meV) and NA (94\,meV) values\,\cite{bib:saitta08,bib:calandra10}.

In order to resolve this discrepancy numerous explanations emerged. A significant temperature dependence of the phonon spectrum lead numerous studies to ascribe the foregoing anomalies to the anharmonicity\,\cite{bib:kortus01,bib:yildirim01,bib:liu01,bib:choi02b,bib:mialitsin07}. Conversely, it was shown that the phonon-phonon corrections constitute only a small portion of the $E_{2g}$ phonon linewidth and the frequency shift, as well as bring about a minor temperature change\,\cite{bib:rafailov02,bib:shukla03,bib:lazzeri03,bib:calandra07,bib:dastuto07}. The effects of the electron relaxation processes (e.g., higher order electron-phonon scattering) on the phonon spectrum were also taken into consideration, since it was shown that high-frequency optical modes in metallic systems might be rather sensitive to these processes when $\mathbf{q}\approx 0$\,\cite{bib:cerdeira72,bib:ipatova74,bib:allen74,bib:kostur91,bib:itai92,bib:marsiglio92,bib:maksimov96}. In fact, few studies have qualitatively demonstrated that it is precisely this mechanism that prompts the breakdown of the standard adiabatic and NA theories in MgB$_2$\,\cite{bib:rafailov02,bib:calandra05,bib:cappelluti06,bib:calandra07,bib:saitta08,bib:ponosov17}. Nevertheless, a concomitant NA theory based in first principles that can resolve this controversy is still absent. Such an in-depth quantitative survey of the NA effects is of great fundamental interest, e.g., for comprehending superconductivity mechanisms, especially in MgB$_2$, where the unusually strong interaction between the $E_{2g}$ phonon and electrons is suspected to underly the electron pairing processes and the ensuing superconductivity state\,\cite{bib:nagamatsu01,bib:kortus01,bib:choi02a}.

Here I present an \emph{ab initio} NA theory for simulating the long wavelength part of the phonon spectrum. The method accounts for the dynamical higher-order electron-phonon scattering processes, whereby the electron-phonon-induced lifetime and energy renormalization of the electron-hole pair excitations are properly treated. The latter two quantities introduce additional energy and temperature dependencies into the spectrum, which is contrary to the common studies\,\cite{bib:saitta08,bib:giustino17}, where only phenomenological lifetimes were taken into account. The results display how such a theoretical description of the NA and relaxation effects is sufficient for reproducing the atypical phonon spectrum of MgB$_2$. Specifically, I show that inclusion of both the electron-hole pair lifetime and energy renormalization mechanisms are crucial for obtaining the final frequency of the $E_{2g}$ phonon. Furthermore, these scattering processes give rise to temperature dependencies of the $E_{2g}$ phonon linewidth and peak position, in close agreement with the experiments. In addition, the electron-phonon vertex corrections entering the phonon spectral function are shown to be negligible, which confirms the validity of the Migdal's theorem\,\cite{bib:migdal58} in MgB$_2$. Finally, I stress that the presented theoretical framework is quite general and can be employed in other interesting cases where electron relaxation processes are expected to be decisive. For instance, it could help elucidate the processes responsible for the large broadening of the $E_{2g}$ phonon in highly doped graphene\,\cite{bib:howard11} or graphite\,\cite{bib:saitta08}, as well as for the anomalous temperature behavior of hexagonal-close-packed transition metals\,\cite{bib:ponosov16}.

Within the many-body perturbation theory, the NA phonon properties are usually extracted from the phonon propagator\,\cite{bib:novko16,bib:giustino17,bib:caruso17},
\begin{eqnarray}
D_{\nu}(\mathbf{q},\omega)=\frac{2\omega_{\mathbf{q}\nu}^{\mathrm{A}}}{\omega^2-(\omega_{\mathbf{q}\nu}^{\mathrm{A}})^2-2\omega_{\mathbf{q}\nu}^{\mathrm{A}}\pi_{\nu}(\mathbf{q},\omega)},
\label{eq:eq1}
\end{eqnarray}
where $\mathbf{q}$ and $\nu$ are the phonon momentum and band index, respectively, $\omega_{\mathbf{q}\nu}^{\mathrm{A}}$ is the adiabatic phonon frequency, and $\pi_{\nu}$ is the phonon self-energy due to the electron-phonon interaction. The real part of $\pi_{\nu}$ then gives the renormalization of $\omega$ due to NA coupling, i.e., $\omega^2=(\omega_{\mathbf{q}\nu}^{\mathrm{A}})^2+2\omega_{\mathbf{q}\nu}^{\mathrm{A}}\mathrm{Re}\,\pi_{\nu}(\mathbf{q},\omega)$, while the imaginary part corresponds to the NA phonon linewidth, i.e.,  $\gamma_{\mathbf{q}\nu}=-2\mathrm{Im}\,\pi_{\nu}(\mathbf{q},\omega)$. In this particular study, I take $\nu=E_{2g}$ of MgB$_2$. The phonon spectral function is obtained by taking the imaginary part of $D_{\nu}$, i.e., $B_{\nu}(\mathbf{q},\omega)=-\pi^{-1}\mathrm{Im}\,D_{\nu}(\mathbf{q},\omega)$. Since Raman spectroscopy probes the long-wavelength part of the phonon spectra, only the $\mathbf{q}\approx0$ limit of $\pi_{\nu}$ and $B_{\nu}$ is to be considered. 

The lowest-order correction over the adiabatic phonon spectral function (i.e., in the case of noninteracting electrons and in the absence of disorders) comes from the dynamic bare interband,
\begin{eqnarray}
\pi_{\nu}^{\mathrm{0,inter}}(\omega)=\sum_{\mu\neq\mu'\mathbf{k}}&&\frac{-\omega\left| g_{\nu}^{\mu\mu'}(\mathbf{k},0) \right|^2}{\varepsilon_{\mu\mathbf{k}}-\varepsilon_{\mu'\mathbf{k}}}\nonumber\\
&&\times\frac{f(\varepsilon_{\mu\mathbf{k}})-f(\varepsilon_{\mu'\mathbf{k}})}{\omega+\varepsilon_{\mu\mathbf{k}}-\varepsilon_{\mu'\mathbf{k}}+i\eta},
\label{eq:eq2}
\end{eqnarray}
and bare intraband,
\begin{eqnarray}
\pi_{\nu}^{\mathrm{0,intra}}(\omega)=\sum_{\mu\mathbf{k}}\left| g_{\nu}^{\mu\mu}(\mathbf{k},0) \right|^2\left[-\frac{\partial f(\varepsilon_{\mu\mathbf{k}})}{\partial\varepsilon_{\mu\mathbf{k}}}\right],
\label{eq:eq3}
\end{eqnarray}
phonon self-energies\,\cite{bib:lazzeri06,bib:saitta08,bib:novko16}. The electron band index and momentum are represented with $\mu$ and $\mathbf{k}$, respectively, $\varepsilon_{\mu\mathbf{k}}$ is the corresponding electron energy, $f(\varepsilon_{\mu\mathbf{k}})$ is the Fermi-Dirac distribution function, and $g_{\nu}^{\mu\mu'}$ is the electron-phonon coupling function. The interband contribution at $\mathbf{q}\approx 0$ was shown to be negligible in MgB$_2$, and, therefore, will not be discussed here further\,\cite{bib:calandra05}. Note that $\pi_{\nu}^{\mathrm{0,intra}}$ is a purely real quantity and thus  only contributes to the renormalization of the phonon frequency. On the other hand, when electron-phonon scattering processes up to all orders are taken into account, the intraband phonon self-energy acquires the following form,
\begin{eqnarray}
\pi_{\nu}^{\mathrm{intra}}(\omega)&=&\sum_{\mu\mathbf{k}}\left| g_{\nu}^{\mu\mu}(\mathbf{k},0) \right|^2\left[-\frac{\partial f(\varepsilon_{\mu\mathbf{k}})}{\partial\varepsilon_{\mu\mathbf{k}}}\right]\nonumber\\
&&\times\frac{\omega}{\omega\left[1+\lambda_{n}(\omega)\right]+i/\tau_{n}(\omega)}.
\label{eq:eq4}
\end{eqnarray}
To reach this result, I adopted a diagrammatic analysis along with the Bethe-Salpeter equation for the phonon self-energy, an approach more commonly utilized for including the dynamical electron-phonon scattering mechanisms into an optical conductivity formula\,\cite{bib:allen71,bib:kupcic07,bib:kupcic09,bib:kupcic15,bib:novko17}. The above expression is equivalent to the intraband phonon self-energy derived in Refs.\,\cite{bib:marsiglio92,bib:maksimov96} by means of Green's functions.
In order to return to the bare nonadiabatic result it is enough to remove the renormalization and scattering time parameters, i.e., $\lambda_n=0$ and $1/\tau_n=0$. The scattering time parameter describes the damping rate (i.e., inverse lifetime) of the excited electron-hole pairs and for the case of electron-phonon scattering can be written as follows\,\cite{bib:allen71,bib:allen74},
\begin{eqnarray}
1/\tau_{n}(\omega)&=&\pi\int d\varepsilon \frac{f(\varepsilon)-f(\varepsilon+\omega)}{\omega}\int d\Omega \alpha^2_{n} F(\Omega)\nonumber\\
&&\times\left[4n_b(\Omega)+f(\Omega+\varepsilon+\omega)+f(\Omega-\varepsilon-\omega)\right.\nonumber\\
&&\left.+f(\Omega+\varepsilon)+f(\Omega-\varepsilon)\right],
\label{eq:eq5}
\end{eqnarray}
where $n_b(\Omega)$ is the Bose-Einstein distribution function and $\alpha^2_{n} F(\Omega)$ is the Eliashberg function with vertex corrections included. In the calculations, I use $\alpha^2_{0} F(\Omega)$ and $\alpha^2_{\mathrm{tr}} F(\Omega)$, where vertex corrections are neglected and included as in the transport theory\,\cite{bib:allen71,bib:giustino17}, respectively\,\footnote{Note that the actual vertex correction term entering the phonon self-energy via Eq.\,\eqref{eq:eq5} has the form $\propto-g_{\nu}(\mathbf{k+q'})/g_{\nu}(\mathbf{k})$\,\cite{bib:novko18}, while the vertex correction term from the transport theory is $\propto-v_{\alpha}(\mathbf{k+q'})/v_{\alpha}(\mathbf{k})$, where $v_{\alpha}(\mathbf{k})$ is the electron group velocity for the polarization direction $\alpha$. In both cases forward scattering events ($\mathbf{q'} \approx 0$) are suppressed.}. Please note that the damping rate depends on the temperature via both the Fermi-Dirac and Bose-Einstein distribution functions. The dynamical renormalization parameter $\lambda_n(\omega)$ is obtained by performing the Kramers-Kronig transformation of $1/\tau_{n}(\omega)$. The static value of the renormalization parameter $\lambda_n(0)$ corresponds to the standard electron-phonon coupling constant.

\begin{figure}[t]
\includegraphics[width=0.5\textwidth]{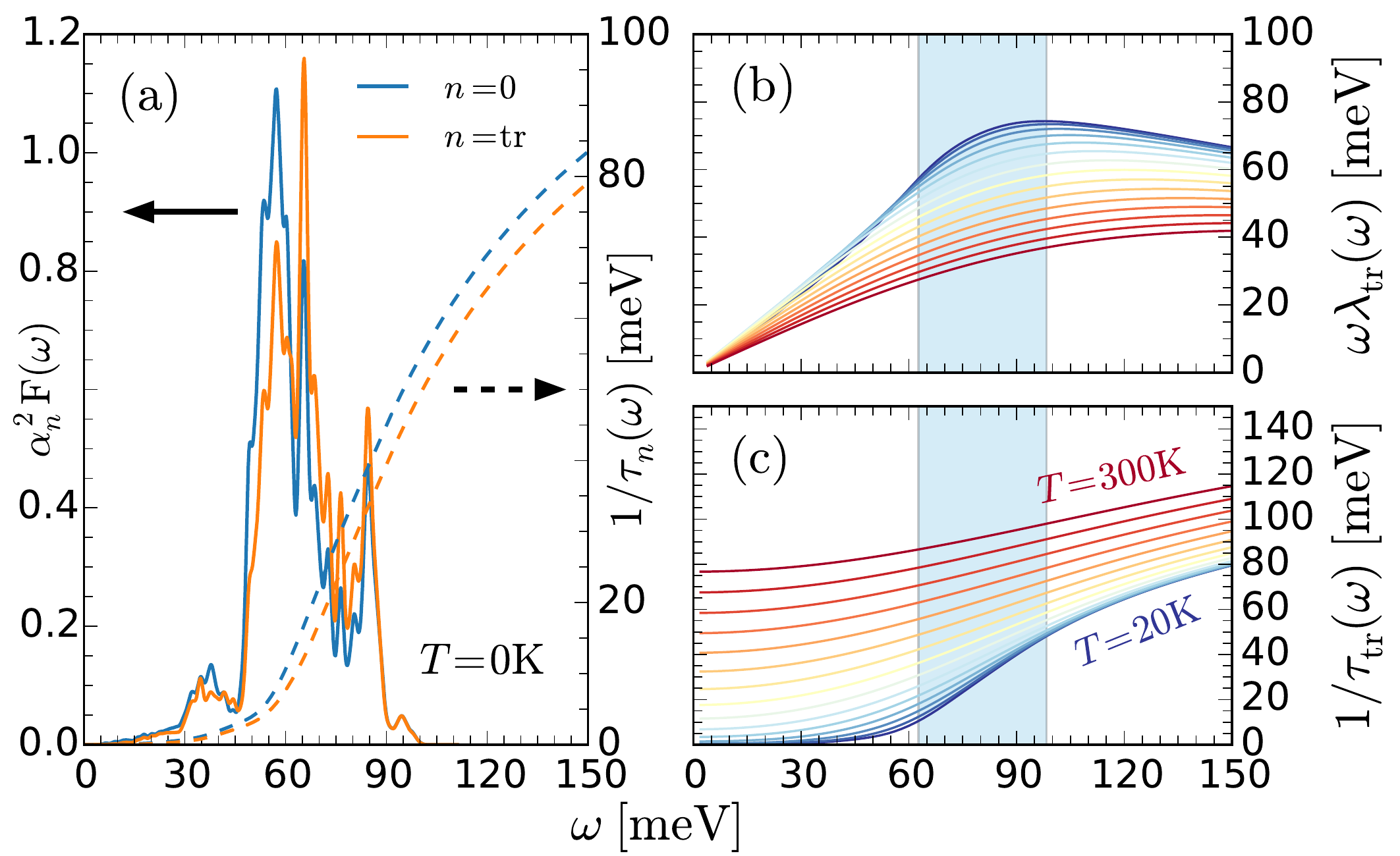}
\caption{\label{fig:fig1}(a) Eliashberg function of MgB$_2$ calculated with and without vertex corrections, i.e., $\alpha^2_{\mathrm{tr}} F(\Omega)$ and $\alpha^2_{0} F(\Omega)$, respectively (left $y$ axis). Resultant damping rates, i.e., $1/\tau_{\mathrm{tr}}(\omega)$ and $1/\tau_{0}(\omega)$, are shown with the dashed lines (right $y$ axis). Temperature dependence of the electron-hole pair (b) energy renormalization parameter $\omega\lambda_{\mathrm{tr}}(\omega)$ and (c) relaxation rate $1/\tau_{\mathrm{tr}}(\omega)$. The blue shaded area represents the window between adiabatic and NA frequencies of the $E_{2g}$ mode.}
\end{figure}

\begin{table}[b]
\caption{\label{tab:table1}Four different electron-phonon scattering regimes and the corresponding phonon frequencies $\omega_{0\nu}$ and linewidths $\gamma_{0\nu}$ valid for a general metallic system with a negligible interband phonon self-energy. The following abbreviations are used: $\alpha=2N(\varepsilon_F)\langle | g_{\nu}^{\mu\mu} |^2 \rangle_{\varepsilon_F}/\omega_{0\nu}^{\mathrm{A}}$, $\alpha^{\ast}=\alpha/(1+\lambda_n)$, and $\tau_n^{\ast}=\tau_n(1+\lambda_n)$, where $\langle \dots \rangle_{\varepsilon_F}$ stands for average over the Fermi surface. Notice that $\pi^{0,\mathrm{intra}}_{\nu}=\alpha\omega^{\mathrm{A}}_{0\nu}/2$.}
\begin{ruledtabular}
\begin{tabular}{clcc}
&\multicolumn{1}{c}{Condition} & $\omega_{0\nu}^2/(\omega^{\mathrm{A}}_{0\nu})^2$ & $\gamma_{0\nu}$\\
\hline
\rule{0pt}{3ex}
(i) & $\lambda_n=0$, $1/\tau_n=0$ & $1+\alpha$ & $0$ \\
(ii) & $\lambda_n=0$, $1/\tau_n\ll\omega$ & $1+\alpha[1-(\omega\tau_n)^{-2}]$ & $\alpha\omega^{\mathrm{A}}_{0\nu}/\omega\tau_n$ \\
(iii) & $\lambda_n=0$, $1/\tau_n\gg\omega$ & $1$ & $\alpha\omega^{\mathrm{A}}_{0\nu}\omega\tau_n$ \\
(iv) & $\lambda_n\neq0$, $1/\tau^{\ast}_n\ll\omega$ & $1+\alpha^{\ast}[1-(\omega\tau^{\ast}_n)^{-2}]$ & $\alpha^{\ast}\omega^{\mathrm{A}}_{0\nu}/\omega\tau^{\ast}_n$ \\
\end{tabular}
\end{ruledtabular}
\end{table}
 
\begin{figure*}[!htbp]
\includegraphics[width=0.95\textwidth]{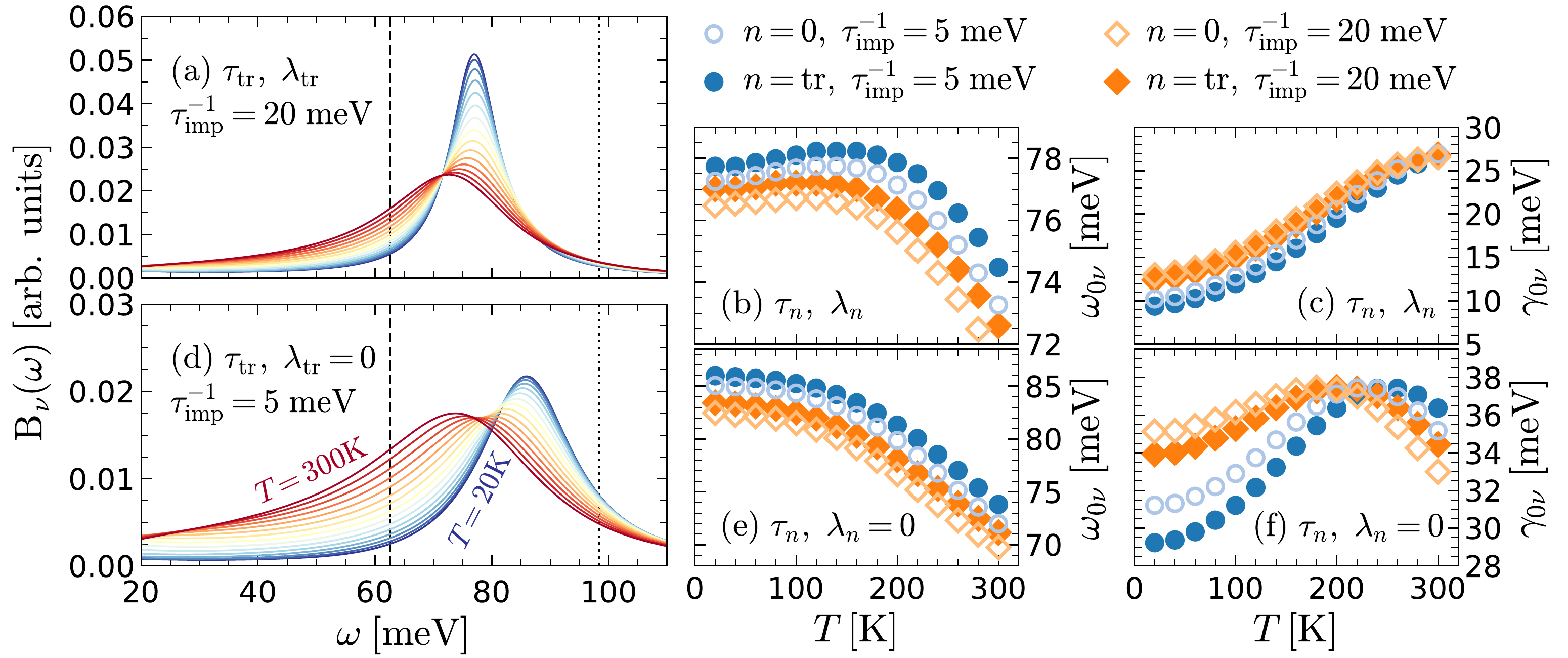}
\caption{\label{fig:fig2} (a) Phonon spectral function of MgB$_2$ as a function of temperature when the electron-hole pair lifetime and energy renormalization effects along with the vertex corrections are included. Dashed and dotted vertical lines represent the obtained adiabatic and NA $E_{2g}$ frequencies, respectively. (b), (c) The corresponding $E_{2g}$ peak positions and linewidths. The plots in (d), (e), and (f) show the same as (a), (b), and (c), but without energy renormalization effects, i.e., $\lambda_n=0$.} 
\end{figure*}
%
%
Figure\,\ref{fig:fig1}(a) displays the results of $\alpha^2_{0} F(\omega)$, $\alpha^2_{\mathrm{tr}} F(\omega)$, and the concomitant relaxation rates, i.e.,  $1/\tau_{0}(\omega)$ and $1/\tau_{\mathrm{tr}}(\omega)$, at $T=0$\,K in MgB$_2$\,\cite{bib:compdetails}. The Eliashberg function $\alpha^2_{0} F(\omega)$ is characterized by the two intense peaks at around 60 and 90\,meV, corresponding to the adiabatic value of the $E_{2g}$ phonon at $\mathbf{q\approx 0}$ and close to the edge of the Brillouin zone\,\cite{bib:shukla03}. The resultant relaxation rate $1/\tau_{0}(\omega)$ shows an abrupt increase at around 60\,meV, reaching quite high values already at 150\,meV. In a previous study, an energy-independent relaxation rate was reported to be $1/\tau=107\,$meV\,\cite{bib:saitta08}, which turns out to be an overestimation according to the results in Fig.\,\ref{fig:fig1}(a).
The obtained electron-phonon coupling constant $\lambda_0(0)$ is 0.78, showing a good agreement with previous \emph{ab initio} calculations\,\cite{bib:eiguren08,bib:calandra10,bib:margine13}. When the vertex corrections (i.e., cancellation of the forward scattering events) are included, the weight of the phonon spectral function is reduced and shifted to slightly higher energies, leading to a less intense damping rate $1/\tau_{\mathrm{tr}}(\omega)$ and coupling constant $\lambda_{\mathrm{tr}}(0)=0.73$.

In Figs.\,\ref{fig:fig1}(b) and \ref{fig:fig1}(c) the temperature dependence of $\omega\lambda_{\mathrm{tr}}(\omega)$ and $1/\tau_{\mathrm{tr}}(\omega)$ is plotted. As the temperature of the system increases, the number of thermally excited phonon modes rises, which in turn results in an augmentation of the electron-phonon scattering probability. This is reflected in the strong enhancement of the damping rate $1/\tau_{\mathrm{tr}}(\omega)$ with temperature, e.g., $1/\tau_{\mathrm{tr}}(0)$ changes from 0 to about 80\,meV when the temperature rises from 20 to 300\,K. The opposite trend is seen for the energy renormalization parameter $\omega\lambda_{\mathrm{tr}}(\omega)$. 

Before discussing the calculated phonon spectral function, I show in Table\,\ref{tab:table1} different electron-phonon scattering regimes along with the resultant phonon frequencies $\omega_{0\nu}$ and linewidths $\gamma_{0\nu}$\,\footnote{Keep in mind that these results are valid for any kind of metallic system, but only when the interband phonon self-energy of the studied mode is insignificant, which holds true for the $E_{2g}$ mode in MgB$_2$.}. (i) When the electron-phonon scattering effects are negligible, the phonon linewidth is zero and the adiabatic frequency is corrected by the real part of the bare intraband phonon self-energy, as it is usually done in the NA studies\,\cite{bib:saitta08,bib:giustino17,bib:caruso17}. (ii) Further, when electron-hole pair relaxation processes are active but small and when $\lambda_n=0$, the phonon linewidth becomes finite, while the phonon frequency is reduced compared to the NA value. (iii) Interestingly, it turns out that strong electron-phonon scattering washes out the NA corrections and the final phonon frequency is close to the adiabatic value\,\cite{bib:maksimov96}. (iv) Finally, the strong electron-hole pair energy renormalization due to electron-phonon coupling along with the finite damping rate reduces the phonon frequency even more than in case (ii). All in all, having in mind the results of $\omega\lambda_{\mathrm{tr}}(\omega)$ and $1/\tau_{\mathrm{tr}}(\omega)$ presented in Figs.\,\ref{fig:fig1}(b) and \ref{fig:fig1}(c), the case of MgB$_2$ falls somewhere between the latter two regimes, i.e., (iii) and (iv), depending on the temperature.

The calculated phonon spectral function, peak position $\omega_{0\nu}$, and linewidth $\gamma_{0\nu}$ of the $E_{2g}$ phonon with electron-hole relaxation processes included are shown in Fig.\,\ref{fig:fig2} as a function of temperature. Along with the electron-phonon, these results include the effects of the electron-impurity scattering in the form of a phenomenological damping rate $1/\tau_{\mathrm{imp}}$ that enters Eq.\,\eqref{eq:eq4} as $1/\tau_n(\omega)\rightarrow 1/\tau_n(\omega)+1/\tau_{\mathrm{imp}}$. The impacts of neglecting the vertex corrections and the energy renormalization parameter $\lambda_n$ are shown as well. Regardless of the applied approximation, all of the presented results show that the electron-phonon scattering processes soften the NA phonon frequency and, thus, improve the agreement with the experiments. As the temperature increases, the $E_{2g}$ peak broadens and loses intensity [see Figs.\,\ref{fig:fig2}(a), \ref{fig:fig2}(c), \ref{fig:fig2}(d), and \ref{fig:fig2}(f)]. At the same time, the frequency slightly increases and after around 200\,K starts to drop [Fig.\,\ref{fig:fig2}(b)] or decreases all the way [Fig.\,\ref{fig:fig2}(e)], depending on whether or not the renormalization parameter $\lambda_n(\omega)$ is included. The inclusion of the vertex corrections blue shifts the frequencies and reduces the linewidths. Nevertheless, the corresponding temperature dependencies are unaltered and the overall effect is quite small, which proves the validity of the Migdal's theorem in the MgB$_2$ phonon spectral function\,\cite{bib:migdal58}. Furthermore, stronger electron-impurity scattering (i.e., larger $1/\tau_{\mathrm{imp}}$) decreases the frequency and increases the linewidth.

Next, I directly compare in Table\,\ref{tab:table2} the obtained peak position at low temperature $\omega^{T\approx 0}_{0\nu}$, as well as the frequency and linewidth change from 20 to 300\,K, i.e., $\Delta\omega^{T}_{0\nu}$ and $\Delta\gamma^{T}_{0\nu}$, with the experiments\,\cite{bib:quilty02,bib:rafailov02,bib:shi04,bib:dastuto07,bib:ponosov17} and a previous (parameter-based) theoretical study\,\cite{bib:cappelluti06}. The results show a remarkable agreement with the experiments for both the low-$T$ peak position and temperature dependencies of $\omega_{0\nu}$ and $\gamma_{0\nu}$. In fact, accounting for the renormalization parameter $\lambda_n(\omega)$ along with the relaxation time $\tau_n(\omega)$ turns out to be essential for obtaining the final accuracy of the peak frequency and the correct temperature behavior. An earlier study reported the value of $\omega^{T\approx 0}_{0\nu}$ in good agreement with the experiments, however, only when a large value of the damping rate was used, i.e., $1/\tau=107\,$meV\,\cite{bib:saitta08}. On the contrary, here I show that a much smaller damping rate due to electron-phonon scattering is needed, i.e., $1/\tau_{n}(\omega^{T\approx 0}_{0\nu})\approx 20$\,meV, when $\lambda_n(\omega)$ is used [see Fig.\,\ref{fig:fig1}(a)]. This is in reasonable agreement with the parameter-based study of Ref.\,\cite{bib:cappelluti06}, where both lifetime and renormalization effects due to electron-phonon coupling were included. The importance of energy renormalization effects is also reflected in the qualitative behavior of $\omega_{0\nu}$ as a function of $T$.
Namely, some experimental measurements reported a constant value or even a small increase of the $E_{2g}$ frequency at lower values of $T$ (i.e., up to about 200\,K)\,\cite{bib:martinho03,bib:mialitsin07,bib:dastuto07,bib:ponosov17}, which is only observed here when $\lambda_n(\omega)$ is included [compare Figs.\,\ref{fig:fig2}(b) and \ref{fig:fig2}(e)].
According to the regime (iv) presented in Table\,\ref{tab:table1}, the insertion of a finite $\lambda_n(\omega)$ decreases the value of $\omega_{0\nu}$ relative to the case (ii), where $\lambda_n(\omega)=0$. Therefore, a slight increase of frequency at lower temperatures is due to the decrease of $\lambda_n(\omega)$ with $T$ [see Fig.\,\ref{fig:fig1}(b)]. Once the $\lambda_n(\omega)$ is low, the temperature dependence of $\omega_{0\nu}$ is then governed solely by $1/\tau_n(\omega)$.

%
%

%
%
%
%

%

\begin{table}[t]
\caption{\label{tab:table2}Frequency of the $E_{2g}$ phonon at low temperature $\omega^{T\approx 0}_{0\nu}$ along with the corresponding frequency $\Delta\omega^{T}_{0\nu}$ and linewidth $\Delta\gamma^{T}_{0\nu}$ shifts from 20 to 300\,K. The results of the NA theory with only lifetime effects ($\tau_n,\lambda_n=0$) as well as with both lifetime and energy renormalization effects ($\tau_n,\lambda_n$) included are shown. The presented ranges include the results obtained with and without the vertex corrections and with $1/\tau_{\mathrm{imp}}=5-20$\,meV. The experimental\,\cite{bib:quilty02,bib:rafailov02,bib:shi04,bib:dastuto07,bib:ponosov17} and previous theoretical\,\cite{bib:cappelluti06} results are shown for comparison. All the reported values are in meV.}
\begin{ruledtabular}
\begin{tabular}{ccccccccc}
 &  \multicolumn{5}{c}{Experiment} & \multicolumn{3}{c}{Theory}\\
\cline{2-6}\cline{7-9}
 & \cite{bib:quilty02} & \cite{bib:rafailov02} & \cite{bib:shi04} & \cite{bib:dastuto07} & \cite{bib:ponosov17} & \cite{bib:cappelluti06} & $\tau_n,\lambda_n=0$ & $\tau_n,\lambda_n$\\ \hline
\rule{0pt}{3ex}
$\omega^{T\approx0}_{0\nu}$ & 77 & 77 &  79 & 81 & 76   & 70 & 82.5-86.0    & 76.5-77.7 \\
$-\Delta\omega^{T}_{0\nu}$  & 2  & 3  &  5  & 7  & $-$1 & 15 & 12.2-13.1    & 3.3-5.0   \\
$\Delta\gamma^{T}_{0\nu}$   & 20 & 9  &  3  & 15 & 17   & 20 & $-$2.1-7.2 & 13.6-17.4 \\
\end{tabular}
\end{ruledtabular}
\end{table}

As a final remark, I note that the presented quantitative and predictive theory can be useful in comprehending the large phonon linewidths of the $E_{2g}$ phonon in highly electron-doped graphene\,\cite{bib:howard11} and graphite\,\cite{bib:saitta08}, since the observed feature is believed to be governed by an unusually strong coupling between the studied phonon mode and electron continuum, producing broad Fano-like shapes in the spectrum. Even more, the above theory accounts for the electron-phonon-induced temperature effects via electron and phonon distribution functions, and, therefore, allows for a resolution in electron and phonon temperatures [see Eq.\,\eqref{eq:eq5}]. This could aid in illuminating microscopic relaxation mechanisms behind the transient phonon frequency shifts obtained in the ultrafast Raman\,\cite{bib:yan09,bib:wu12,bib:ferrante18} and optical\,\cite{bib:ishioka08} spectroscopies, where a nonequilibrium condition between electrons and phonons is achieved.

In summary, I have presented an accurate nonadiabatic theoretical framework based in first principles, in which the effects of the electron-hole pair relaxation due to dynamical coupling with phonons are taken into account. These processes enter the intraband phonon self-energy through the energy- and temperature-dependent lifetime and renormalization parameters. By using this methodology, I have simulated the temperature-dependent linewidth and frequency of the $E_{2g}$ phonon in MgB$_2$ in close agreement with the experiments. Lifetime and energy renormalization effects due to electron-phonon scattering were both found to be decisive in this regard. Stated differently, electron-mediated coupling between the $E_{2g}$ phonon and the rest of the energetically available phonon modes of MgB$_2$ were proven to be responsible for the unusual, temperature-dependent spectral features.

\begin{acknowledgments}
The author gratefully acknowledges financial support from the European Regional Development Fund for the ``Center of Excellence for Advanced Materials and Sensing Devices'' (Grant No. KK.01.1.1.01.0001). Financial support by Donostia International Physics Center (DIPC) during various stages of this work is also highly acknowledged. Computational resources were provided by the DIPC computing center.
\end{acknowledgments}

\newpage
\bibliography{namgb}

\begin{thebibliography}{62}
\bibitem [{bib()}]{bib:compdetails}%
  \BibitemOpen
  \href@noop {} {}\bibinfo {note} {The ground-state calculations were done by
  means of the {\sc quantum espresso} package\,\cite{bib:qe} with a plane-wave
  cutoff energy of 40 Ry. Norm-conserving pseudopotentials were used with the
  Perdew-Zunger version of the local density approximation (LDA)
  exchange-correlation functional\,\cite{bib:lda}. A ($12\times12\times12$)
  Monkhorst-Pack grid was used for sampling the Brillouin zone. Adiabatic
  phonon frequencies were calculated on a ($6\times6\times6$) grid using
  density functional perturbation theory\,\cite{bib:baroni01}. All the phonon
  self-energies [Eqs.\,\eqref{eq:eq2}-\eqref{eq:eq4}] were calculated with the
  personally modified version of the {\sc epw} code\,\cite{bib:epw}, while the
  corresponding input quantities (electron energies, phonon frequencies, and
  electron-phonon coupling elements) were interpolated using maximally
  localized Wannier functions\,\cite{bib:wannier}. The $\mathbf{k}$ summations
  in the phonon self-energies were done on a ($300\times300\times300$) grid.
  The Eliashberg functions needed for calculating the relaxation rate
  Eq.\,\eqref{eq:eq5} were done on $(30\times30\times30)$ electron and
  $(20\times20\times20)$ phonon momentum grids.}\BibitemShut {Stop}%
\end{thebibliography}

\end{document}